\documentclass[aps,prd,twocolumn,nofootinbib,superscriptaddress]{revtex4}
\usepackage{amsmath} 
\usepackage{verbatim} 
\usepackage{mathrsfs}
\usepackage{amsfonts}
\usepackage{amssymb}
\usepackage{graphicx}
\usepackage{color}
\usepackage{hyperref}

\usepackage{soul}

\usepackage[utf8]{inputenc}
% ...
\inputencoding{latin1}
% ...
\inputencoding{utf8}

\begin{document}

\title{Comment on ``Cosmic Microwave Background Constraints Cast a Shadow On Continuous Spontaneous
Localization Models'' }

\author{Gabriel R. Bengochea}
\email{gabriel@iafe.uba.ar}
\affiliation{Instituto de Astronom\'\i
	a y F\'\i sica del Espacio (IAFE), CONICET--Universidad de Buenos Aires, 1428 Buenos Aires, Argentina}

\author{Gabriel Le\'{o}n}
\email{gleon@fcaglp.unlp.edu.ar }
\affiliation{Grupo de Astrof\'{\i}sica, Relatividad y Cosmolog\'{\i}a, Facultad
	de Ciencias Astron\'{o}micas y Geof\'{\i}sicas, Universidad Nacional de La
	Plata, Paseo del Bosque S/N 1900 La Plata, Buenos Aires, Argentina}
\affiliation{CONICET, Godoy Cruz 2290, 1425 Ciudad Aut\'onoma de Buenos Aires, Argentina. }

\author{Philip Pearle}
\email{ppearle@hamilton.edu}
\affiliation{Emeritus, Department of Physics, Hamilton College, Clinton, NY 13323, USA}

\author{Daniel Sudarsky}
\email{sudarsky@nucleares.unam.mx}
\affiliation{Instituto de Ciencias Nucleares, Universidad
	Nacional Aut\'{o}noma de M\'{e}xico, A.P. 70-543, M\'{e}xico
	D.F. 04510, M\'{e}xico.}
\begin{abstract}
In a recent paper [J. Martin \& V. Vennin, Phys. Rev. Lett. 124, 080402 (2020)] it was argued that, for most natural choices, the direct application of the continuous spontaneous localization (CSL) theory to the inflationary case, as it is known to work in non-relativistic laboratory situations, is ruled out by cosmological observational data, thus casting a shadow on models based on CSL theory. We point out that such results are based on the consideration of a rather narrow set of choices for the application of the theory to the cosmological context and that the landscape of open and different possibilities is extremely vast.
\end{abstract}

\maketitle

The main point in this comment is that when attempting the extrapolation required to take a theory designed to work in non-relativistic laboratory conditions, with negligible  space-time curvature, to regimes where matter is characterized by relativistic quantum fields (QF) living in a highly curved spacetime (as in the case of an exponentially expanding universe), one must confront several non-trivial issues and the landscape of open possibilities is extremely large. Thus, when viewed in such perspective, the shadow alluded to in Ref. \cite{jmartinPRL} must be recognized as a rather small one.

We illustrate some of the issues one must face, starting by recalling the basic elements appearing in the non-relativistic CSL theory and how its parameters are introduced. Appearing in the state vector evolution equation are $\lambda$ and $\hat A({\bf x})$. The parameter $\lambda$ is the collapse rate for a neutron in a spatially superposed state. And
\begin{equation}\label{smeared}
\hat A({\bf x})=C\frac{m}{M_{N}}\int d{\bf z}e^{-[{\bf x}-\bf{z}]^{2}/4r_{c}^{2}}\hat N({\bf z}),
\end{equation}
where $\hat N({\bf z})=\hat \xi^{\dagger}({\bf z})\hat \xi({\bf z})$ is the particle number density operator, constructed from (non-relativistic) creation and annihilation operators,
$m$ is the mass of the corresponding particle species ($M_N$ is the mass of the neutron), where such mass-proportionality is suggested by empirical evidence \cite{Pearle1994}, and the smearing length $r_{c}$ is the second parameter of the theory. $\hat A({\bf x})$, the collapse-generating operator (toward one of whose eigenstates the collapsing state vector tends), may be thought of as a \textit{smeared mass density} operator at ${\bf x}$. The choices $\lambda\approx 10^{-16}$s$^{-1}$and $r_{c}\approx 10^{-5}$cm, originating in the SL collapse theory \cite{GRW}, were provisionally adopted by Pearle \cite{Pearle1989} for CSL as providing sound behavior when applied to laboratory situations.

In quantum field theory in curved spacetimes (QFTCS), the notion of particle disappears at the fundamental level \cite{Wald94}, leaving only the quantum fields themselves.
Therefore, the notion of ``localization'', when considering a quantum field, is quite different than that pertaining to a particle. That is, the main quantum uncertainties characterizing the state of a particle are connected with its position in space (and its momentum), making the appearance of the fundamental length $r_c$ natural. The uncertainties characterizing the state of a quantum field are instead connected with the {\it ``value of the field''} at each point (and the conjugate momentum), so the corresponding natural parameter must have the same dimensions as the corresponding field, call it $\Delta$. The relation that might exist between the parameter $r_c$, and $\Delta$ is then far from clear (more on this below). A similar problem arises when considering the collapse rate parameter $\lambda$, characterizing phenomena at laboratory scales. In relating the two regimes (laboratory experiments and the early inflationary era) it must be recognized that the universe passes through several important phase transitions, including nucleosynthesis, hadronization, the electro-weak transition, and more uncertain regimes such as reheating and the end of inflation. Thus, there are ample grounds to doubt any simple connection between the values of the parameters $r_c$ and $\lambda$, relevant for one regime, with the parameters characterizing the theory in a completely different one. The need for a relativistic version of CSL is in fact briefly acknowledged in \cite{jmartinPRL}.

On the other hand, when considering a suitable collapse operator for the cosmological epoch ruled by the inflaton field, one might seek something that reduces to the mass density operator in the laboratory setting. This takes us to consider the energy-momentum tensor $T_{ab}$. However, this choice possesses several issues, e.g. the question of smearing becomes more complex, as it can be expected to combine some smearing in field space and some smearing of space-time. Should one extract a scalar operator (such as  $T_a^a$ or $(T_{ab}T^{ab})^{1/2}$, etc)? or somehow work with the full tensor $T_{ab}$?

Furthermore, the CSL parameters might not be fundamental constants. They could possibly depend on quantities characterizing the environment, just as the masses in the standard model of particles depend on a vacuum expectation value in the Higgs sector. This naturally suggests a possible effective time dependence of the CSL parameters in the cosmological context, since the values of relevant fields such as the inflaton and the Higgs field, or even spacetime curvature (a possibility one can tie to the idea of a fundamental  connection of collapse and gravitation \cite{diosi1989, penrose1996}), experienced dramatic changes between the inflationary epoch and the present day. In \cite{jmartinPRL}, the possibility of a standard QFT type of running of the CSL parameters is mentioned, but the point we want to emphasize is that an effective time dependence can result from a much wider set of possibilities.

Moreover, one must find a way to combine QFTCS including CSL type modifications with GR. Here we note the simplest options: i) a full semi-classical approach as developed in \cite{alberto}, and ii) the traditional approach (which is only defined at the perturbative level) where one quantizes metric and scalar field perturbations (such as the Mukhanov-Sasaki variable). The last one is the route taken by \cite{jmartinPRL}, along with many cosmologists, but it should be noted that there are problematic aspects in that framework. For instance, the requirement that field commutation relations follow the full spacetime causal structure are not strictly enforced (one just relies on the causal structure of the background rather than that of the complete metric) e.g. \cite{Causal2, Causal1}. Of course the first path also faces complex difficulties \cite{benito}.

In all these regards, the authors of \cite{jmartinPRL} make some specific choices. For instance they take the collapse-generating operator to be something they identify as {\it the density contrast operator}, an object whose particular form depends on adopting i) or ii) mentioned above. The authors perform the smearing of such an operator as if one was dealing with the non-relativistic theory Eq. \eqref{smeared}, including parameter values as well, and encounter incompatibilities for these parameter values in the inflationary setting. The point we note is that such choices are far from unique, the theory landscape is vast, and in fact we have provided several arguments to cast doubts on whether those choices are {\it``the natural ones"}. Finally, and as acknowledged by those authors themselves, without some kind of modification to the standard Quantum Theory, such as that offered by CSL, the account of the emergence of the seeds of cosmic structure during inflation suffers from very serious shortcomings \cite{Shortcomings}. The task ahead is therefore the exploration of the theoretical landscape broadly depicted here, to find a formulation of the theory that reduces to non-relativistic CSL (so far satisfactory for laboratory contexts). Such an extrapolation must be generally covariant, suitable for QFTCS applications and empirically successful in the inflationary context.

\begin{acknowledgments}
We thank J. Martin \& V. Vennin for their very helpful comments on a previous daft. G.R.B. is supported by CONICET (Argentina) and he acknowledges support from grant PIP 112-2017-0100220CO of CONICET (Argentina). G.L. is supported by CONICET (Argentina) and the National Agency for the Promotion of Science and Technology (ANPCYT) of Argentina grant PICT-2016-0081. D.S. acknowledges partial financial support from  PAPIIT-UNAM, Mexico (Grant No.IG100120); the Foundational Questions Institute (Grant No. FQXi-MGB-1928); the Fetzer Franklin Fund, a donor advised by the Silicon Valley Community Foundation.

\end{acknowledgments}

\bibliography{bibliografia}
\bibliographystyle{apsrev}

\end{document}